\documentclass[12pt]{article}
\usepackage{amsmath}
\usepackage{graphicx}
\usepackage{amsfonts}
\usepackage{amssymb}
\newtheorem{theorem}{Theorem}[section]

\newtheorem{lemma}[theorem]{Lemma}

\newtheorem{remark}[theorem]{Remark}
\newenvironment{proof}[1][Proof]{\textsc{#1.} }{\ \rule{0.5em}{0.5em}}
\numberwithin{equation}{section}

\begin{document}

\title{Einstein - Bianchi system with sources}
\author{Yvonne Choquet-Bruhat and James W. York}
\maketitle
\begin{abstract}
We write a first order symmetric hyperbolic system coupling the Riemann tensor
with the derivative of an electromagnetic 2 - form and the dynamical
acceleration of a perfect relativistic fluid. We determine the associated,
coupled, Bel - Robinson type energy, and the integral equality that it satisfies.
\end{abstract}

\textit{Dedicated to Guy Boillat with deep esteem and affection.}

\section{Introduction.}

The interest of formulating fundamental physical laws as first order symmetric
hyperbolic systems has been stressed, after the foundational work of
K.\ O.\ Friedrichs, by various authors. Particularly important developments in
the case of systems of conservation laws have been obtained by Lax, Boillat,
Ruggeri and Strumia, Anile, Dafermos. They have used the existence, in many
physical cases, of an additional convex conserved density, associated to an
entropy functional. They have obtained symmetric first order hyperbolic
systems in many cases of physical interest, by introducing auxiliary variables
similar to Lagrange multipliers.

In the case of Einstein's gravitation theory, the problem seems more delicate.
There is no local density of gravitational energy, nor entropy. The effective
strength of the gravitational field lies in the Riemann tensor of the
spacetime metric. Its evolution is governed by the so - called higher order
equations (Bel 1958, Lichnerowicz 1964), deduced from the Bianchi identities.
The system satisfied by the trace free part of the Riemann tensor, the Weyl
tensor, was some time ago recognized as a linear, first order symmetrizable
hyperbolic system (FOSH), with constraints, homogeneous in vacuum. See H.
Friedrich (1996) and references therein. The evolution equations for the
Riemann tensor itself, now renamed \textit{Bianchi equations}, have also been
written (CB-Yo 1997, 2001) as a FOSH\ system, made explicit in terms of four
two- tensors, introduced by Bel (1958), the electric and magnetic
gravitational fields and corresponding inductions relative to a Cauchy adapted
frame. A FOSH system has also been written for the Bianchi equations (no
longer homogeneous) and the dynamical acceleration of a perfect fluid
(CB-Yo2002). In this article, dedicated to one of the best specialist of these
FOSH systems, we couple the Bianchi equations with the equations satisfied by
the dynamical acceleration of a charged fluid and the derivatives of the
associated Maxwell field.

\section{Einstein equations with sources.}

\subsection{Definitions.}

The Einstein equations, on a 4 - dimensional manifold $V,$ link the Ricci
tensor of a pseudo Riemannian metric $g$ of Lorentzian signature with the
stress energy tensor $T$ of sources. These equationsread
\begin{equation}
Ricci(g)=\rho\equiv T-\frac{1}{2}g(tr_{g}T),
\end{equation}
that is, in local coordinates,
\[
R_{\alpha\beta}=\rho_{\alpha\beta},\text{ \ with \ }\rho_{\alpha\beta}\equiv
T_{\alpha\beta}-\frac{1}{2}g_{\alpha\beta}T_{\lambda}^{\lambda}.
\]

The stress energy tensor of an electrically charged, relativistic, perfect
fluid is the sum of the fluid stress energy tensor:
\[
t_{\alpha\beta}\equiv(\mu+p)u_{\alpha}u_{\beta}+pg_{\alpha\beta}%
\]
and the Maxwell tensor:
\begin{equation}
\tau^{\alpha\beta}\equiv F^{\alpha}{}_{\lambda}F^{\beta\lambda}-\frac{1}%
{4}g^{a\beta}F^{\lambda\mu}F_{\lambda\mu}.
\end{equation}
In these formulas $u^{\alpha}$ is the unit kinematical velocity satisfying
$u^{\alpha}u_{\alpha}=-1;$ $\mu,p,S$ are the specific energy, pressure and
entropy. These thermodynamic quantities are assumed to be positive and linked
by an equation of state:
\begin{equation}
\mu=\mu(p,S).
\end{equation}
The entropy $S$ satisfies a conservation equation\footnote{{\footnotesize Such
an equation can be deduced as usual from a particle number density
conservation and the Gibbs relation.}} along the flow lines:
\begin{equation}
u^{\alpha}\partial_{\alpha}S=0.
\end{equation}
The tensor $F$ is the antisymmetric electromagnetic closed 2 - form. It
satisfies the Maxwell equations:
\begin{equation}
dF=0,\text{ \ and \ }\delta F=J,
\end{equation}
where $d$ and $\delta$ are respectively the differential and codifferential
operator (in the metric $g),$ while $J$ is the electric current. We suppose
$J$ to be a convection current, that is we have:
\begin{equation}
J=qu,\text{ \ i.e. \ }J^{\alpha}=qu^{\alpha},
\end{equation}
with $q$ the electric charge density.

\subsection{Fluid equations.}

The conservation laws, consequence of the Einstein equations, read:
\begin{equation}
\nabla_{\alpha}T^{\alpha\beta}\equiv\nabla_{\alpha}(t^{\alpha\beta}%
+\tau^{\alpha\beta})=0.
\end{equation}

The Maxwell equations imply that, with $J^{\lambda}F^{\beta}{}_{\lambda}$
called the Lorentz force,
\begin{equation}
\nabla_{\alpha}\tau^{\alpha\beta}=J^{\lambda}F^{\beta}{}_{\lambda}.
\end{equation}

The index $f$ of the fluid, is given by the identity
\begin{equation}
f(p,S):=\exp\int_{p_{0}}^{p}\frac{dp}{\mu(p,S)+p},
\end{equation}

We introduce the \textbf{dynamical velocity, }tangent vector to the flow
lines, defined by
\begin{equation}
C^{\alpha}:=fu^{\alpha},\text{ \ \ hence \ \ \ }C^{\alpha}C_{\alpha}=-f^{2}.
\end{equation}
The dynamical velocity incorporates information on the kinematic velocity
$u^{\alpha}$ and the thermodynamic quantities. The definitions of $C$ and $f$
imply that:
\begin{equation}
-C^{\beta}\nabla_{\alpha}C_{\beta}\equiv f\frac{\partial f}{\partial
x^{\alpha}}=f^{2}\frac{\partial_{\alpha}p}{\mu+p}+f\frac{\partial f}{\partial
S}\partial_{\alpha}S.
\end{equation}
The entropy equation 2.4 and the tensor $t_{\alpha\beta}$ become:
\begin{equation}
C^{\alpha}\partial_{\alpha}S=0,
\end{equation}%
\begin{equation}
t_{\alpha\beta}\equiv(\mu+p)f^{-2}C_{\alpha}C_{\beta}+g_{\alpha\beta}p
\end{equation}
Therefore the equations 2.7 can be written:
\begin{equation}
C^{\beta}\nabla_{\alpha}[(\mu+p)f^{-2}C^{\alpha}]+(\mu+p)f^{-2}C^{\alpha
}\nabla_{\alpha}C^{\beta}+\partial^{\beta}p=-J^{\lambda}F^{\beta}{}_{\lambda}.
\end{equation}

We contract with $C_{\beta}$ these equations and we use the equation 2.11 and
the entropy conservation 2.12 to obtain the following continuity equation
(under our hypothesis on $J$, the Lorentz force $J^{\lambda}F^{\beta}%
{}_{\lambda}$ is orthogonal to $C)$
\begin{equation}
\nabla_{\alpha}[(\mu+p)f^{-2}C^{\alpha}]=0,
\end{equation}
which can be written:
\begin{equation}
\nabla_{\alpha}C^{\alpha}+(\mu_{p}^{\prime}-1)\frac{C^{\alpha}C^{\beta}%
}{C^{\lambda}C_{\lambda}}\nabla_{\alpha}C_{\beta}=0,
\end{equation}
where $\mu_{p}^{\prime}:=\frac{\partial\mu}{\partial p}$ is a given function
of $p$ and $S.$

One deduces from 2.14 and 2.15 the following equations of motion
\begin{equation}
C^{\alpha}\nabla_{\alpha}C^{\beta}+f^{2}\frac{\partial^{\beta}p}{\mu+p}%
=-\frac{f^{2}}{\mu+p}J^{\lambda}F^{\beta}{}_{\lambda}%
\end{equation}
Using 2.11 we see that the equations of motion may be written:
\begin{equation}
C^{\alpha}\{\nabla_{\alpha}C_{\beta}-\nabla_{\beta}C_{\alpha}+(\mu
+p)^{-1}(-C^{\lambda}C_{\lambda})^{\frac{1}{2}}qF_{\beta\alpha}\}+\frac{1}%
{2}\frac{\partial f^{2}}{\partial S}\partial_{\beta}S=0.
\end{equation}

\begin{remark}
These equations are not independent, their left hand side is orthogonal to $C.$
\end{remark}

In these equations the fluid unknowns are the four components of the vector
$C^{\alpha},$ and the scalar $S.$ The specific pressure $p$ is a known
function of $f$ (i.e. of $C^{\alpha},$ by 2.10) and of $S$%
\begin{equation}
p\equiv p(C^{\alpha}C_{\alpha},S),
\end{equation}
determined by inverting the relation 2.3.

We suppose that $S$ is initially constant, it remains then constant during the
evolution and the term $\partial_{\alpha}S$ disappears from the equations.\ It
can be proved, by adapting the Rendall
symmetrization\footnote{{\footnotesize General techniques, using an additional
convex entropy function, of Lax 1957, Boillat 1994, Ruggeri and Strumia 1981
(reported in the book by Anile 1982), could also be used.}} of the perfect
uncharged fluid, that the equations 2.16 and 2.17\ can be written as a first
order symmetric hyperbolic system for the dynamical velocity $C.$ The
electromagnetic field $F$ appears as underivated in these equations.

\section{Bianchi equations.}

In contradistinction with the fluid and electromagnetic sources, the writing
of the Einstein equations 2.1\ as a first order system for the derivatives of
the metric requires a choice of gauge. It was done straightforwardly by Fisher
and Marsden using harmonic (now called wave) coordinates. In these coordinates
the Ricci tensor takes the form of a quasidiagonal, quasilinear system of wave
equations for the metric $g.$ Though the non linear stability of Minkowski
spacetime has now been proved (Lindblad and Rodnianski 2005) using these
coordinates, the fundamental geometrical object signaling the existence of a
gravitational field remains the Riemann tensor.

The Bianchi equations satisfied by the Riemann tensor of a metric $g,$
solution of the Einstein equations 2.1, are:%

\begin{equation}
\nabla_{0}R_{hi,\lambda\mu}+\nabla_{i}R_{0h,\lambda\mu}-\nabla_{h}%
R_{0i,\lambda\mu}=0
\end{equation}
and
\begin{equation}
-\nabla_{0}R^{0}{}_{i,\lambda\mu}-\nabla_{h}R_{:::i,\lambda\mu}^{h}%
=-J_{\lambda,0\mu}\equiv-\nabla_{0}\rho_{\mu\lambda}+\nabla_{\mu}%
\rho_{0\lambda}%
\end{equation}

We recall that a sliced Lorentzian manifold $(V,g)$ is a product $M\times R$
endowed with a Lorentzian metric $g$ which induces on each $M_{t}%
:=M\times\{t\}$ a Riemannian metric $\bar{g}_{t}.$ The metric $g$ takes in a
Cauchy adapted moving frame, i.e \ a frame with its timelike axis orthogonal
to the space slices, the usual 3+1 form
\begin{equation}
ds^{2}=-N^{2}dt^{2}+g_{ij}(\bar{\theta}^{i}+\beta^{i}dt)(\bar{\theta}%
^{j}+\beta^{j}dt).
\end{equation}
The $g_{ij}$ are the components of $\bar{g}_{t}$ in a coframe $\bar{\theta
}^{i}$ on $M.$ Particular cases are the natural coframe $\bar{\theta}%
^{i}=dx^{i},$ and an orthonormal coframe $\bar{\theta}^{i}=a_{j}^{i}dx^{j}%
\ $where $g_{ij}=\delta_{ij}.$ We denote by $g^{hk}$ the contravariant
components of $\bar{g}_{t},$ they are, in our spacetime frames, also the $hk$
contravariant components of $g.$ The derivatives $\partial_{\alpha}$ are the
Pfaff derivatives in the coframe $\theta^{0}=dt,$ \ \ $\theta^{i}=\bar{\theta
}^{i}+\beta^{i}dt,$ that is
\begin{equation}
\partial_{0}=\frac{\partial}{\partial t}-\beta^{i}\partial_{i},\text{
\ \ }\partial_{i}=\bar{\partial}_{i}=A_{i}^{j}\frac{\partial}{\partial x^{j}},
\end{equation}
with $A$ the inverse matrix of $a$ whose elements are $a_{j}^{i}.$

\begin{theorem}
The equations 3.1 and 3.2. are a first order symmetric hyperbolic system for
the components of the Riemann tensor in a Cauchy adapted frame, with
$\bar{\theta}^{i}$ an orthonormal frame.
\end{theorem}

\begin{proof}
The principal operator of these equations is diagonal by blocks, a block
corresponding to the derivatives of the components $R_{hi,\lambda\mu}$ and
$R_{0h,\lambda\mu.}$ for a given pair $\lambda,\mu,$ $\lambda<\mu.$ Such a
block is the following symmetric matrix, if $\bar{\theta}^{i}$ is an
orthonormal coframe :

$\ \ \ \ \ \ \ \ \ \ \ \ \ \ \ \ \ \ \ \ \ \ \ \ \ \ \ \ \ \ \ \ \ \ \left(
\begin{tabular}
[c]{cccccc}%
$N^{-2}\partial_{0}$ & 0 & 0 & $\partial_{2}$ & $-\partial_{1}$ & 0\\
0 & $N^{-2}\partial_{0}$ & 0 & 0 & $\partial_{3}$ & $-\partial_{2}$\\
0 & 0 & $N^{-2}\partial_{0}$ & $-\partial_{3}$ & 0 & $\partial_{1}$\\
$\partial_{2}$ & 0 & $-\partial_{3}$ & $\partial_{0}$ & 0 & 0\\
$-\partial_{1}$ & $\partial_{3}$ & 0 & 0 & $\partial_{0}$ & 0\\
0 & $-\partial_{2}$ & $\partial_{1}$ & 0 & 0 & $\partial_{0}$%
\end{tabular}
\right)  $.

Each block is symmetric, hence the full principal operator is also symmetric.

The matrix $M^{t}$ of the coefficients of derivatives $\frac{\partial
}{\partial t}$ is identical to the matrix $M^{0}$ of the coefficients of the
derivatives $\partial_{0}.$ It is diagonal, with coefficients either $1$ or
$N^{-2},$ it is therefore positive definite.
\end{proof}

The energy on a $t=$constant submanifold $M_{t},$ constant $t,$ associated to
the Bianchi equations is the so - called Bel Robinson energy, integral on
$M_{t}$ of the positive definite quadratic form $M^{t}$ in the Riemann tensor.
If one introduces the electric and magnetic fields and inductions space 2-
tensors associated with the Riemann tensor\footnote{$\eta_{ijk}$
{\footnotesize the volume form on M}$_{t}.${\footnotesize .}} given by
\[
E_{ij}\equiv R^{0}{}_{i,0j},\text{ \ }D_{ij}\equiv\frac{1}{4}\eta_{ihk}%
\eta_{jlm}R^{hk,lm},\text{ }%
\]%
\[
\text{\ }H_{ij}\equiv\frac{1}{2}N^{-1}\eta_{ihk}R^{hk},_{oj},\text{ \ }%
B_{ji}\equiv\frac{1}{2}N^{-1}\eta_{ihk}R_{0j},^{hk},
\]
the Bel - Robinson energy density on $M_{t}$ is given by:
\[
\mathcal{E}_{g}\equiv\frac{1}{2}(|E|^{2}+|H|^{2}+|D|^{2}+|B|^{2}).
\]

In the case that we are considering of equations with sources, the Bianchi
equations contain, in addition to the Riemann tensor and its gradient, the
dynamical acceleration $\nabla C$ and the gradient $\nabla F$ of $F,$ but no
derivative of these quantities.

\section{Equations for $\nabla F.$}

The Maxwell equations imply
\begin{equation}
\delta dF+d\delta F=dJ
\end{equation}
that is, using the Ricci identity and some manipulation of indices, the
following semilinear wave equation for $F:$%
\begin{equation}
\nabla^{\alpha}\nabla_{\alpha}F_{\beta\gamma}=f_{\beta\gamma},
\end{equation}
with
\begin{equation}
f_{\beta\gamma}:=R_{\beta}{}^{\lambda}F_{\gamma\lambda}-R_{\gamma}{}^{\lambda
}F_{\lambda\beta}-2R_{\beta}{}^{\lambda},_{\gamma}{}^{\alpha}F_{\lambda\alpha
}+\nabla_{\beta}J_{\gamma}-\nabla_{\gamma}J_{\beta}.
\end{equation}
It is easy to deduce a symmetric hyperbolic first order system for $\nabla F$
from the wave equation satisfied by $F.$ We set
\begin{equation}
F_{\gamma,\alpha\beta}:=\nabla_{\gamma}F_{\alpha\beta},
\end{equation}
In a Cauchy adapted frame with orthonormal $\bar{\theta}^{i},$ the equations
to satisfy read:
\begin{equation}
N^{-2}\nabla_{0}F_{0,\alpha\beta}-\delta^{ij}\nabla_{i}F_{j,\alpha\beta
}=f_{\alpha\beta}%
\end{equation}
and
\begin{equation}
\nabla_{0}F_{i,\alpha\beta}-\nabla_{i}F_{0,\alpha\beta}=R_{0i,\alpha}%
{}^{\lambda}F_{\lambda\beta}+R_{0i,\beta}{}^{\lambda}F_{\alpha\lambda}.
\end{equation}
The left hand sides of these equations are a symmetric hyperbolic first order
operator for the four unknowns ($F_{0,\alpha\beta},F_{i,\alpha\beta})$ for
each pair ($\alpha,\beta),$ $\alpha<\beta.$ The right hand sides contain
$\nabla C$ and $Riemann(g),$ but no derivative of these quantities.

The energy density associated to the above system is a positive quadratic
form, $\mathcal{E}_{\nabla F}$ in $\nabla F.$ It is sometimes called a ''superenergy''.

\section{Equations for $\nabla C.$}

The dynamical acceleration $\nabla C$ satisfies the following equations
obtained by covariant differentiation of 2.16. and 2.17, and use of the Ricci
identities:\
\begin{equation}
M_{\gamma\beta}\equiv C^{\alpha}(\nabla_{\alpha}C_{\gamma\beta}-\nabla_{\beta
}C_{\gamma\alpha})+a_{\gamma\beta}=0
\end{equation}
and
\begin{equation}
g^{\alpha\beta}\nabla_{\alpha}C_{\gamma\beta}+(\mu_{p}^{\prime}-1)\frac
{C^{\alpha}C^{\beta}}{C^{\lambda}C_{\lambda}}\nabla_{\alpha}C_{\gamma\beta
}+b_{\gamma}=0
\end{equation}
where we have set
\begin{equation}
C_{\gamma\beta}\equiv\nabla_{\gamma}C_{\beta},
\end{equation}%
\begin{equation}
a_{\gamma\beta}\equiv C_{\gamma}{}^{\alpha}(C_{\alpha\beta}-C_{\beta\alpha
})+C^{\alpha}C_{\lambda}R_{\gamma\alpha,\beta}^{.........\lambda}%
+\nabla_{\gamma}\{(\mu+p)^{-1}(-C^{\lambda}C_{\lambda})^{\frac{1}{2}%
}qC^{\alpha}F_{\beta\alpha}\},
\end{equation}%
\begin{equation}
b_{\gamma}\equiv-R_{\gamma\lambda}C^{\lambda}+\nabla_{\gamma}\{(\mu
_{p}^{\prime}-1)\frac{C^{\alpha}C^{\beta}}{C^{\lambda}C_{\lambda}}%
\}C_{\alpha\beta}%
\end{equation}
The last term in $b_{\gamma}$ is a quadratic form in $C_{\alpha\beta}$ whose
coefficients are functions of the $C^{\alpha}$ and $S.$ These functions can be
computed by using the identity (we use $S=$ constant)
\[
\nabla_{\gamma}\mu_{p}^{\prime}\equiv\mu_{p^{2}}^{\prime\prime}\partial
_{\gamma}p.
\]
By the definition of $f$ and the identity 2.4 it holds that
\[
\partial_{\gamma}p=(\mu+p)f^{-1}\partial_{\gamma}f=-(\mu+p)(C^{\lambda
}C_{\lambda})^{-1}C^{\alpha}C_{\gamma\alpha}.
\]

The equations 5.1. are not independent, because they satisfy the identities
\begin{equation}
C^{\beta}M_{\gamma\beta}\equiv0,
\end{equation}
the equations 5.1 and 5.2. are not a well posed system. Instead of the
$4\times4$ equations 4.1 we consider\footnote{{\footnotesize An analogous
procedure is used for the symmetrization of the Euler equations in K.O.
Friedrichs 1969\ and in Rendall 1992.}} the $4\times3$ ones:
\begin{equation}
\tilde{M}_{\gamma i}\equiv M_{\gamma i}-\frac{C_{i}}{C_{0}}M_{\gamma0}=0
\end{equation}
The terms in derivatives of $C_{\gamma\lambda}$ in these equations can be
written in the following form:
\begin{equation}
C^{\alpha}\partial_{\alpha}(C_{\gamma i}-\frac{C_{i}}{C_{0}}C_{\gamma
0})-(\partial_{i}-\frac{C_{i}}{C_{0}}\partial_{0})(C^{\alpha}C_{\gamma\alpha})
\end{equation}

\begin{lemma}
The system 5.2,5.7. is equivalent to a FOS (First Order Symmetric) system
\ for $C_{\gamma\alpha}$ with right hand side function of the Riemann tensor
and $\nabla F,$ not of \ their derivatives.
\end{lemma}

\begin{proof}
The system is quasi diagonal by blocks, each block corresponding to a given
value of the index $\gamma.$ We will write the principal operator of a block
by omitting this index. We set
\begin{equation}
U_{i}\equiv C_{\gamma i}-\frac{C_{i}}{C_{0}}C_{\gamma0},\text{ \ \ \ }%
U_{0}\equiv C^{\alpha}C_{\gamma\alpha}%
\end{equation}
and we define the differential operators $\tilde{\partial}_{\alpha}$ as
follows:
\begin{equation}
\tilde{\partial}_{0}\equiv C^{\alpha}\partial_{\alpha},\text{ \ \ \ }%
\tilde{\partial}_{i}=\partial_{i}-\frac{C_{i}}{C_{0}}\partial_{0}%
\end{equation}
The principal terms (derivatives of $C_{\gamma\alpha})$ in the equations 5.7
with index $\gamma$ are
\begin{equation}
\tilde{\partial}_{0}U_{i}-\tilde{\partial}_{i}U_{0}.
\end{equation}
We have by inverting 5.9:
\[
C_{\gamma0}\equiv\frac{C_{0}(U_{0}-C^{i}U_{i})}{C^{\lambda}C_{\lambda}}%
\]%
\[
C_{\gamma i}\equiv U_{i}+\frac{C_{i}(U_{0}-C^{j}U_{j})}{C^{\lambda}C_{\lambda
}}%
\]
The principal terms of 5.2. read, using the above formulae
\begin{equation}
\frac{\mu_{p}^{\prime}C^{\alpha}\partial_{\alpha}U_{0}}{C^{\lambda}C_{\lambda
}}+(g^{ij}-\frac{C^{i}C^{j}}{C^{\lambda}C_{\lambda}})\partial_{i}U_{j}%
-\frac{C^{0}C^{i}}{C^{\lambda}C_{\lambda}}\partial_{0}U_{i}%
\end{equation}
We introduce the positive definite (if $C$ is timelike) quadratic form
\begin{equation}
\tilde{g}^{ij}\equiv g^{ij}-\frac{C^{i}C^{j}}{C^{\lambda}C_{\lambda}}.
\end{equation}
Then we find that
\[
\tilde{g}^{ij}\frac{C_{j}}{C_{0}}\equiv\frac{C^{0}C^{i}}{C^{\lambda}%
C_{\lambda}}%
\]
The principal terms 5.12 are therefore
\begin{equation}
\frac{\mu_{p}^{\prime}\tilde{\partial}_{0}U_{0}}{C^{\lambda}C_{\lambda}%
}+\tilde{g}^{ij}\tilde{\partial}_{i}U_{j}%
\end{equation}
The matrix of the coefficients of the derivatives $\tilde{\partial}_{\alpha}$
in the equations deduced from the system 5.2, 5.7 is

$\ \ \ \ \ \ \ \ \ \ \ \ \ \ \ \ \ \ \ \ \ \ \ \ \ \ \ \ \ \ \left(
\begin{array}
[c]{cccc}%
-\frac{\mu_{p}^{\prime}}{C^{\lambda}C_{\lambda}}\tilde{\partial}_{0} &
-\tilde{\partial}^{1} & -\tilde{\partial}^{2} & -\tilde{\partial}^{3}\\
-\tilde{\partial}_{1} & \tilde{\partial}_{0} & 0 & 0\\
-\tilde{\partial}_{2} & 0 & \tilde{\partial}_{0} & 0\\
-\tilde{\partial}_{3} & 0 & 0 & \tilde{\partial}_{0}%
\end{array}
\right)  $

This matrix is symmetrized by taking the product with the $4\times4$ matrix

$\ \ \ \ \ \ \ \ \ \ \ \ \ \ \ \ \ \ \ \ \ \ \ \ \ \ \ \ \ \ \ \ \ \ \ \ \ \ \ \ \ \ \ \ \left(
\begin{array}
[c]{cc}%
1 & 0\\
0 & \tilde{g}_{ij}%
\end{array}
\right)  .$
\end{proof}

\textit{Hyperbolicity}.

In the case we are considering the matrix $\tilde{M}^{0}$ is diagonal, with
positive elements if $\mu_{p}^{\prime}>0$ and $C^{\alpha}$ is timelike. The
principal matrix written with natural coordinates is also symmetrizable. The
corresponding matrix $M^{t}$ is not diagonal, and it is not obvious that it is
positive definite. In fact it will be so only if $\mu_{p}^{\prime}\geq1$. We
will prove the following lemma.

\begin{lemma}
The system 5.2, 5.7 is FOSH if $\mu_{p}^{\prime}\geq1$ and $C$ is timelike.
\end{lemma}

\begin{proof}
It is simpler to compute directly the energy density for the considered
system: its positivity is equivalent to the positivity of the matrix $M^{t}$.
Multiplying 5.2. and 5.7 respectively by $U_{0}$ and $\tilde{g}^{ij}U_{j}$
gives equations of the form
\begin{equation}
\frac{1}{2}\frac{\mu_{p}^{\prime}\tilde{\partial}_{0}(U_{0})^{2}}{f^{2}%
}-\tilde{g}^{ij}U_{0}\tilde{\partial}_{i}U_{j}=U_{0}\Phi_{0}%
\end{equation}%
\begin{equation}
\tilde{g}^{ij}U_{j}\tilde{\partial}_{0}U_{i}-\tilde{g}^{ij}U_{j}%
\tilde{\partial}_{i}U_{0}=\tilde{g}^{ij}U_{j}\Phi_{i}%
\end{equation}
where the $\Phi_{\alpha}$ contain only non differentiated terms in $U,$
Riemann and $\nabla F$. We add these two equations, replace the operators
$\tilde{\partial}$ by the operators $\partial$ and carry out some
manipulations using the expression for $\tilde{g}^{ij}$ and the Leibniz rule.
We obtain that
\begin{equation}
\partial_{0}\mathcal{E}_{m}+\bar{\nabla}_{i}\mathcal{H}_{m}^{i}=Q_{m}%
\end{equation}
The function $Q_{m}$ is a quadratic form in $C_{\gamma\alpha},$ $F_{\alpha
,\beta\gamma}$ and the Riemann tensor, while $\mathcal{E}_{m}$ is the energy
density on $M_{t}$ of the dynamical acceleration $\nabla C.$ It is the
quadratic form given by:
\begin{equation}
\mathcal{E}_{m}\equiv f^{-2}[(\mu_{p}^{\prime}-1)U_{0}^{2}+(U_{0}-C^{i}%
U_{i})^{2}]+g^{ij}U_{i}U_{j}%
\end{equation}
It is positive definite if $\mu_{p}^{\prime}\geq1$ and $C^{\alpha}$ is timelike.
\end{proof}

\begin{remark}
The system is hyperbolic in the sense of Leray if $\mu_{p}^{\prime}>0,$ but
the submanifolds $x^{0}=$constant are 'spacelike' with respect to the fluid
wave cone only if the fluid sound speed is less than the speed of light, i.e.,
$\mu_{p}^{\prime}\geq1.$
\end{remark}

The quantity $\mathcal{E}_{m}$\ is called the fluid ''acceleration energy'' density.

\section{Coupled system.}

The previous results give, for $Rieman(g),$ $\nabla F$ and $\nabla C,$ when
$g,$ a lorentzian metric, $F,$ a 2 form and $C,$ a timelike vector, are known,
a first order symmetric system, quasi diagonal by blocks, hyperbolic if
$\mu_{p}^{\prime}\geq1$.

By choosing a densitized lapse (time wave gauge), as in [CB-Yo97] and
[CB-Yo01] one can obtain a full symmetric first order system containing, in
addition, $g$ and its connection, $F$ and $C.$

The superenergy density $\mathcal{E}$ for the full system is the sum of the
Bel - Robinson energy density of the gravitational field, the energy density
of $\nabla F,$ supernergy of $F,$ and the energy density of the dynamical
acceleration:
\[
\mathcal{E\equiv E}_{g}\mathcal{+E}_{F}+\mathcal{E}_{m}.
\]
Using the expression of $\partial_{0}$ and the mean extrinsic curvature
$\tau\equiv g^{ij}K_{ij}$ of the space slices $S_{t},$ whose volume element we
denote by $\mu_{\bar{g}_{t}},$ we obtain by integration an integral equality
whose right hand side couples all these superenergies:
\[
\int_{S_{t}}\mathcal{E\mu}_{\bar{g}}=\int_{S_{t_{0}}}\mathcal{E\mu}_{\bar{g}%
}+\int_{t_{0}}^{t}\int_{S_{\theta}}\{-N\tau\mathcal{E}+\mathcal{Q}%
\}\mu_{\tilde{g}}dt
\]
where $\mathcal{Q}$ is a quadratic form in $Riemann(g),$ $\nabla F$ and
$\nabla C$, which could be estimated in terms of $\mathcal{E}$ if the other
unknowns, $C,$ $g$ and its connection, were estimated.

\textbf{References.}

[1] Anile M. 1982 Relativistic fluids and magneto fluids Camb. Univ. press.

[2] Bel L.1958, C.R. Acad. Sciences Paris.246 , 3105-3107.

[3] Boillat G. 1994\ Non linear hyperbolic fields and waves, in ''Recent
Mathematical Methods in non linear Wave propagation, Boillat, Dafermos, Lax
and Liu ed. LN in Math 1640, Springer.

[4] Boillat G. and Ruggeri T 1999 J.\ Math. Phys. 40 6399.

[5] Choquet-Bruhat Y 1966, Comm Math Phys 3, 334-357

[6] Foures (Choquet) - Bruhat Y. 1958 Bull Soc Math France 86, p. 155-175.

[7] Choquet-Bruhat Y, York J, 1997, Banach cent. pub. 41-1\ 119-131.

[8] Choquet - Bruhat Y. and York J. W. 2001 Top Met. in Non lin. An.

[9] Choquet - Bruhat Y. and York J. W.\ 2002 Comptes Rendus Ac. Sciences,
Paris 335 8 711-716.

[10] Dafermos C. Hyperbolic conservation laws in continuum physics 2000, Springer.

[11] Friedrich H. 1996 Class. Quantum Grav 13, 1451-1459.

[12] Friedrich H. 1998, Phys Rev D 57, 2317-2322.

[13] Friedrichs K.O. 1969, Fluid and magnetofluids, Colloque CNRS 1969

[14] Lax P. 1957 Hyperbolic systems of conservation laws. Comm. Pures And App.
Maths 10 537-566.

[13] Leray J. 1953 Hyperbolic differential equations I.A.S, Princeton

[15] Lichnerowicz A. 1964, Annales IHES n$%
{{}^\circ}%
10.$

[16] Rendall A. 1992 J. Math Phys 33, 1047-1053.

[17] Ruggeri T, 2003 Proceedings WASCOM, Monaco R., Pennisi S. Rionero S. and
Ruggeri T. ed. World Scientific, 441-454.

[18] Ruggeri T, Strumia A, Ann. 1981, Ann. H. Poincare 34, 65-84

\bigskip

YCB. Acad. Sciences 23 Quai Conti, Paris 6, France. YCB@ccr.jussieu.fr

JWY. Physics Department, Cornell University, Ithaca, NY, 14853-6801, USA. York@astro.cornell.edu
\end{document}